\documentclass[aps,pre,twocolumn,showpacs,superscriptaddress,eqsecnum]{revtex4}
\usepackage{amsmath,graphicx,amssymb}

\begin{document}

\title{Faraday instability on viscous ferrofluids in a horizontal magnetic field: Oblique rolls
of arbitrary orientation}
\author{V. V. Mekhonoshin}
\affiliation{Institut f\"ur Theoretische Physik, Universit\"at Magdeburg, Universit\"atsplatz 2, D-39106, Magdeburg, Germany}
\affiliation{Institute of Continuous Media Mechanics UB RAS, 1, Korolyev St., 641013, Perm, Russia}
\author{Adrian Lange}
\email[]{Adrian.Lange@physik.uni-magdeburg.de}
\homepage[]{http://itp.nat.uni-magdeburg.de/~adlange}
\affiliation{Institut f\"ur Theoretische Physik, Universit\"at Magdeburg, Universit\"atsplatz 2, D-39106, Magdeburg, Germany}

\date{\today}

\begin{abstract}
A linear stability analysis of the free surface of a horizontally unbounded ferrofluid layer of arbitrary depth 
subjected to vertical vibrations and a horizontal magnetic field is performed. A nonmonotonic dependence of 
the stability threshold on the magnetic field is found at high frequencies of the vibrations. The reasons of the decrease of the 
critical acceleration amplitude caused by a horizontal magnetic field are discussed. It is revealed that the magnetic field can 
be used to select the first unstable pattern of Faraday waves. In particular, a rhombic pattern as a superposition of two
different oblique rolls can occur. A scaling law is presented which maps all data into one graph for the tested range of
viscosities, frequencies, magnetic fields and layer thicknesses.
\end{abstract}

\maketitle

\section{\label{intro} Introduction}

The Faraday instability denotes the parametric generation of standing waves on the free surface of a fluid subjected to vertical 
vibrations. The study of this phenomenon dates back to the observations by Faraday in 1831 \cite{Faraday}. The initially flat free surface of
the fluid becomes unstable at a 
certain intensity of the vertical vibrations of the whole system. As a result of the instability, a pattern of standing waves is formed at 
the fluid surface. The typical response is subharmonic, i.e.,  the wave frequency is half the frequency of the excitation.
The harmonic response can be observed on a shallow fluid at low frequencies \cite{Muller1}. Faraday waves allows
one to investigate symmetry breaking phenomena in a spatially extended nonlinear system. Therefore they experience a renewed interest in 
recent years. A detailed experimental study of the various patterns on a viscous fluid has been performed by Edwards and Fauve 
\cite{Fauve} who used a one-frequency as well as a two-frequency forcing.  Parallel rolls, hexagons, and a twelvefold quasi-pattern were 
observed. Though typical Faraday waves are subharmonic, the twelvefold quasi-pattern appears to be harmonic in a certain region of 
parameters. Binks {\it et al.}  have shown experimentally that the depth of the layer \cite{Binks} and the excitation frequency 
\cite{Binks1} affect significantly the regions, where either rolls or hexagons or squares are observed. In 
\cite{Muller93,Kudrolli,Arbell1,Arbell} it is revealed that a two-frequency excitation leads to a great variety of wave patterns. In 
particular, the observed  patterns were triangles \cite{Muller93}, superlattices formed by small and large hexagons \cite{Kudrolli}, 
squares \cite{Muller93,Kudrolli,Arbell1,Arbell}, and rhomboid pattern \cite{Arbell}. 

The comprehensive linear stability analysis of the Faraday instability on an arbitrarily deep layer of a viscous nonmagnetic fluid has 
been performed by Kumar and Tuckerman \cite{Kumar}. This analysis was tested experimentally in \cite{Bechhoefer} and  an excellent agreement 
between the  predicted and experimental data was found. Weizhong and Rongjue \cite{Weizhong} extended the linear analysis \cite{Kumar} for
the case of an arbitrary periodic excitation. In \cite{Muller1,Cerda} the low frequency region is studied in  great detail. 
Bicritical points, where transitions from one type of response to others occur, are predicted and experimentally confirmed in \cite{Muller1}.
In \cite{Tuckerman,Kumar1} an analogy between the Faraday instability and a periodically driven version of the Rayleigh-Taylor instability is 
exploited. In \cite{Tuckerman} a scaling law is suggested, which satisfactorily describes the behavior of the system in a wide range of 
parameters. Kumar \cite{Kumar1} discusses the mechanism of the wave number selection in the Faraday instability on high-viscous fluids. 

In order to solve the pattern selection problem for the Faraday instability it is necessary to take into account nonlinear interactions 
between different excited modes. The nonlinear behavior of the system has been studied theoretically in a great number of papers (see
\cite{Milner,Zhang,Chen1,Chen,Silber} and references therein). An amplitude equation for an infinitely deep layer of a viscous fluid
is derived by Chen and  Vi\~nals \cite{Chen}. A good agreement between the predicted frequencies for the transition between regions of
different symmetry \cite{Chen} and the experimental observations \cite{Binks1} is found. 

Magnetic fluids (or ferrofluids) are colloidal dispersions of single domain nanoparticles in a carrier liquid. The attractiveness of ferrofluids
stems from the combination of a normal liquid behavior with the sensitivity to magnetic fields. This enables the use of magnetic fields
to control the flow  of the fluid, giving rise to a great variety of new phenomena and to numerous technical applications
\cite{Ferrohydrodynamics}.

One of the most interesting phenomena of pattern formation in ferrofluids is the Rosensweig instability \cite{Rosensweig}.
At a certain intensity of the normal magnetic field the initially flat surface of a horizontal ferrofluid layer becomes unstable.
Peaks appear at the fluid surface, which typically form a static hexagonal pattern at the final stage of the pattern forming
process \cite{Abou00}. By including vertical vibrations to that setup, M\"uller \cite{Muller} analyzed the Faraday instability on viscous
ferrofluids subjected to a vertical magnetic field in the frame of a linear stability analysis. It has been found that the joint action
of the two destabilizing factors leads to a delay of the Rosensweig instability. By an appropriate choice of the system parameters one
can observe either a normal or anomalous dispersion. The predictions of \cite{Muller} were confirmed experimentally in~\cite{Petrelis}. 

Parametric waves on the surface of a ferrofluid can be excited using a vertical \cite{Bacri_EPJ} or a horizontal \cite{Cebers,Bacri} 
alternating magnetic field. This phenomenon is called the magnetic Faraday instability. In \cite{Bacri_EPJ} the dispersion relation 
was investigated experimentally and it displays two significant features. The response of the surface waves is harmonic with respect
to the frequency of the magnetic field and the wave vector of the resulting rolls is parallel to the field, i.e. the crests and
troughs of the rolls are perpendicular to the field. In \cite{Bacri} a supercritical transition from rolls
to rectangles was observed and explained by means of a weakly-nonlinear analysis. 

In the present paper the stability of the surface of a ferrofluid subjected to vertical vibrations and a static horizontal magnetic field 
is studied in a wide range of the system parameters. The horizontal magnetic field tends to decrease the curvature of the ferrofluid 
surface along the direction of the field \cite{Zelazo}, i.e. it tends to stabilize the flat surface. On the other hand, vertical vibrations
tends to destabilize a flat surface. Thus the aim of the linear stability analysis is to study the behavior of 
a ferrofluid subjected to two, in a certain sense competing factors.
It will be shown that the magnetic field allows one to control the stability of the surface significantly and to affect the symmetry of
the linearly most unstable pattern. This opportunity can be used in numerous 
technical applications, where magnetic fields tangential to the fluid surface and perpendicular vibrations are typical. 

\section{\label{system}System and basic equations}

\begin{figure}[bp]
    \includegraphics[scale=0.45]{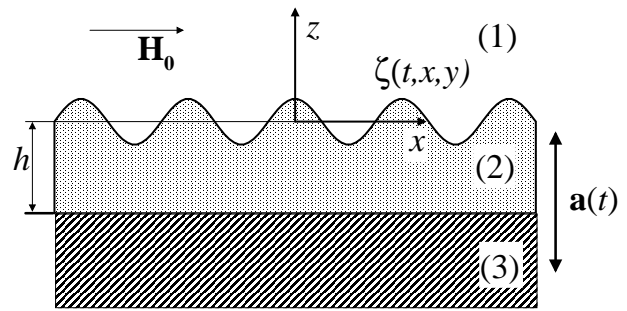}
\caption{A horizontally unbounded ferrofluid layer (2) is  placed in a nonmagnetic container (3) with air 
(1) above. The system is subjected to a horizontal magnetic field $\mathbf{H}_0$ and harmonic vertical vibrations $\mathbf{a}(t)$.}
  \label{setup}
\end{figure}
A dielectric, viscous, and incompressible magnetic fluid with constant density $\rho$ and permeability $\mu_r$ is 
considered. The laterally infinite ferrofluid layer of arbitrary depth $h$ is subjected to a homogeneous dc horizontal magnetic field 
and harmonic vertical vibrations (Fig.\ \ref{setup}). The plane $z=0$ coincides with 
the nondeformed surface of the ferrofluid. The fluid layer is bounded from below by the bottom of the nonmagnetic container and has a free 
surface described by $\zeta(t,x,y)$ with air above. A homogeneous magnetic field 
$\mathbf{H_0}=(H_0,0,0)$ is applied along the $x$-axis. Due to zero electrical conductivity of the 
fluid, the static form of the Maxwell equations is used for the strength $\mathbf{H}$ and the induction $\mathbf{B}$ of the magnetic 
field in all 
three media
\begin{equation}
\text{div}\, \mathbf{B}^{(i)}=0,\qquad\text{rot}\,  \mathbf{H}^{(i)} =0,\qquad i=1, 2,  3\; ,
\label{Maxwell}
\end{equation}
\noindent where superscripts denote the different media: 1 air, 2 magnetic fluid, and 3 container.  
A linear relation between the magnetization of the fluid $\mathbf{M}^{(2)}$ and the 
strength of the magnetic field inside is assumed, $\mathbf{M}^{(2)}=\mu_0(\mu_r-1)\mathbf{H}^{(2)}$.

The fluid motion is governed by the continuity equation and the Navier-Stokes equations 
\begin{subequations}
\label{gov}
\begin{align}
\text{div}\, \mathbf{v} &= 0\; ,\label{cont}\\
\frac{\partial \mathbf{v}}{\partial t}+\left(\mathbf{v}\,\text{grad} \right)  \mathbf{v} &= -\frac 1 \rho\,\text{grad}\, 
p+\nu\Delta \mathbf{v}+\mathbf{g}(t)\; .
\label{Navier}
\end{align}
\end{subequations}

\noindent Here $\mathbf{v}=(u,v,w)$ is the fluid velocity, $p$ is the pressure, and $\nu$ is the kinematic viscosity of the fluid. The 
vertical vibrations add a periodic  term to the gravity acceleration $ 
\mathbf{g}_0$, i.e., a 
modulated value $ \mathbf{g}(t) = (0,0,-g_0-a\,\cos(\omega t))$ appears in the Navier-Stokes equations. Here $a$ is the 
acceleration amplitude and $\omega$ is the angular frequency of the vibrations. The components of the stress tensor 
$\tensor{\mathrm{T}}^{(2)}$ 
read
\begin{equation}
\begin{split}
T_{ij}^{(2)}=&\left\lbrace-p-\mu_0\int\limits_0^H{MdH'}-\mu_0\frac{H^2}{2}\right\rbrace\delta_{ij}\\&+H_iB_j+\rho\nu(\partial_iv_j+
\partial_jv_i) \; .
\end{split}
\end{equation}

The governing set of equations has to be supplemented by the boundary conditions. For the magnetic field the conditions are the decay of 
all 
perturbations far from the ferrofluid ($z\rightarrow\pm\infty$) and the continuity of the normal (tangential) component of the induction 
(strength) of the magnetic field across the air-fluid interface ($z=\zeta$) and at the bottom of the container ($z=-h$)

\begin{subequations}
\label{bound_field}
\begin{alignat}{2}
\mathbf{H}^{(1)}&=\mathbf{H_0}&\quad  \text{at}\quad z&\rightarrow \infty \; ,\label{above}\\
B_n^{(1)}&=B_n^{(2)},\quad  H_{\tau}^{(1)}=H_{\tau}^{(2)}&\quad  \text{at}\quad z&=\zeta \; ,\label{air_mag}\\
B_n^{(2)}&=B_n^{(3)},\quad H_\tau^{(2)}=H_\tau^{(3)}&\quad\text{at}\quad z&=-h \; ,\label{bottom_mag}\\
\mathbf{H}^{(3)}&= \mathbf{H_0}&\quad  \text{at}\quad z  &\rightarrow  -\infty \; .\label{below}
\end{alignat}
\end{subequations}

\noindent The subscripts $n$ and $\tau$ denote the normal and tangential components of the vector.

The hydrodynamic equations are closed by the no-slip condition at the bottom of the container,

\begin{equation}
 \mathbf{v}=0\quad \text{at}\ z=-h\label{no-slip}\; ,
\end{equation}

\noindent and the kinematic boundary condition at the free surface of the fluid

\begin{equation}
w=\partial_t\zeta+( \mathbf{v}\,\text{grad})\zeta\quad \text{at}\ z=\zeta \; .\label{kin}
\end{equation}

\noindent The equations for magnetic field and the fluid motion are coupled by the continuity condition for the stress tensor 
across the air-fluid interface, which completes the statement of the 
problem,

\begin{equation}
\begin{split}
n_j\Bigg\lbrace -p^{(1)}+p+\mu_0\int\limits_0^H{MdH'}+\mu_0\frac{M_n^2}{2}\Bigg\rbrace&\\ -\rho\nu 
n_i(\partial_iv_j+\partial_jv_i)&=\sigma Kn_j \; .
\label{stress}\end{split}
\end{equation}

\noindent Here $\sigma$ is the surface tension, $K=\,\text{div}\, \mathbf{n}$ is the surface curvature, $p^{(1)}$ is the 
atmospheric 
pressure above the 
fluid (assumed to be constant), and $ \mathbf{n}$ is the unit vector normal to the surface given by

\begin{equation}
 \mathbf{n} = \frac{\text{grad}[z-\zeta(t,x,y)]}{|\text{grad} [z-\zeta(t,x,y)]|} \; .
\end{equation} 

\section{\label{Floquet}Linear stability analysis}

Following the standard procedure~\cite{Kumar,Muller}, the governing equations and the boundary conditions have been linearized in the 
vicinity 
of the 
nonperturbed state 
\begin{align*}
\mathbf{v}&=0,\quad \zeta=0,\quad \mathbf{H}^{(i)}=\mathbf{H}_0,\quad i=1, 2, 3 \; ,\\
p_0&=p^{(1)}-\frac{\mu_0}{2}M_0H_0-g(t)z \; , 
\end{align*}

\noindent where $M_0=(\mu_r-1)H_0$ and $p_0$ is the pressure in the unperturbed state.
The linearized governing equations for the small
perturbations, which encompass the magnetic field strength $\mathbf{H}_1$, the pressure $p_1$, and a nonzero velocity $\mathbf{v}$
of the fluid, read
\begin{subequations}
\label{gov_lin}
\begin{align}
\mu_0\mu_r\,\text{div}\, \mathbf{H}_1^{(i)}&=0,\qquad\text{rot}\,  \mathbf{H}_1^{(i)} =0,\quad i=1, 2,  3 \; ,
\label{Maxwell_lin}\\
\text{div}\, \mathbf{v} &= 0 \; ,\label{cont_lin}\\
\frac{\partial \mathbf{v}}{\partial t}&= -\frac 1 \rho\,\text{grad}\,p_1+\nu\Delta \mathbf{v} \; .
\label{Navier_lin}
\end{align}
\end{subequations}
\noindent Here the linear relation between the induction and the strength of the magnetic field is used.  At the first order with respect to 
the perturbations, the boundary conditions are
\begin{subequations}
\label{bound_lin}
\begin{align}
\mathbf{H}_1^{(1)}&=0\quad \text{at}\; z\rightarrow \infty \; ,\label{above_lin}\\
H_{1n}^{(1)}=\mu_rH_{1n}^{(2)},\quad  H_{1\tau}^{(1)}&=H_{1\tau}^{(2)}\quad\text{at}\; z=0 \; ,\label{air_mag_lin}\\
\mu_rH_{1n}^{(2)}=H_{1n}^{(3)},\quad H_{1\tau}^{(2)}&=H_{1\tau}^{(3)}\quad\text{at}\; z=-h \; ,\label{bottom_mag_lin}\\
\mathbf{H}_1^{(3)}&=0\quad\text{at}\; z \rightarrow  -\infty \; ,\label{below_lin}\\
 \mathbf{v}&=0\quad \text{at}\; z=-h \; ,\label{no-slip_lin}\\
w-\partial_t\zeta&=0\quad\text{at}\; z=0 \; ,\label{kin_lin}\\
\partial_zu+\partial_xw&=0\quad\text{at}\; z=0 \; ,\label{strx}\\
\partial_zv+\partial_yw&=0 \quad\text{at}\; z=0 \; ,\label{stry}\\
g(t)\zeta-p_1-\mu_0M_0H_1&+2\rho\nu\partial_zw=\sigma\Delta_\perp\zeta\quad\text{at}\; z=0 \; ,\label{strz}
\end{align}
\end{subequations}

\noindent where $\Delta_\perp=\partial_{xx}+\partial_{yy}$. The stability of the flat surface with 
respect to standing waves is analyzed by using the Floquet ansatz for the surface 
deformations and the $z$-component of the velocity

\begin{subequations}
\label{floquet}
\begin{eqnarray}
\zeta(t,x,y)&=&\sin( \mathbf{kr}) e^{(s+i\alpha\omega)t}\sum\limits_{n=-\infty}^{\infty}{\zeta_n e^{in\omega 
t}} \; ,\qquad\qquad\label{profile}\\
w(t,x,y,z)&=&\sin( \mathbf{kr}) e^{(s+i\alpha\omega)t}\sum\limits_{n=-\infty}^{\infty}{w_n(z) e^{in\omega 
t}}\label{velocity} \; ,
\end{eqnarray}
\end{subequations}

\noindent where $ \mathbf{k} =(k_x,k_y)$ is the two-dimensional wave vector,  $s$ is the growth rate, and $\alpha$ is the parameter 
determining the type of the response. For $\alpha=0$ the response is harmonic whereas for $\alpha=1/2$ it is subharmonic. Expansions 
similar to (\ref{velocity}) are made for all other small 
perturbations and are inserted into the linearized governing equations (\ref{gov_lin}). The functions of the 
vertical coordinate in the Floquet expansion are given by  linear combinations of the exponential functions $e^{\pm kz}$ and $e^{\pm 
q_nz}$ with $q_n^2=k^2+[s+i(\alpha+n)\omega]/\nu$ and $\Re(q_n)>0$. The 
condition of reality for $\zeta(t,x,y)$ leads to the equations \cite{Kumar} 

\begin{subequations}
\label{reality}
\begin{alignat}{2}
\zeta_{-n}&=\zeta_n^{*},&\quad \alpha&=0 \; , \label{real_har}\\
\zeta_{-n}&=\zeta_{n-1}^*,& \quad \alpha&=1/2 \; . \label{real_sub}
\end{alignat}
\end{subequations}

\noindent The boundary conditions (\ref{bound_lin}) allow one to express all the perturbed quantities 
in terms of the coefficients $\zeta_n$ which satisfy the equation

\begin{equation}
\sum\limits_{n=-\infty}^{\infty} {\left( W_{n}\zeta_n-a\zeta_{n-1}-a\zeta_{n+1}\right)e^{[s+i(\alpha+n)\omega ]t}=0} \; ,
\label{disp}
\end{equation}

\noindent where

\begin{equation}
\begin{split}
W_n=&-2 \Bigg[\frac{\nu^2}{k[q_n\,\coth(q_nh)-k\,\coth(kh)]}\\
&\times\Big(q_n[4k^4+(k^2+q_n^2)^2]\,\coth(kh)\,\coth(q_nh)\\&-k[4k^2q_n^2+(k^2+q_n^2)^2]-\frac{4q_nk^2(k^2+q_n^2)}{\sinh(kh)\,\sinh(q_nh)
}\Big)\\ 
&+g_0+\frac{\sigma k^2}{\rho}+\frac{k(\mu_r-1)^2\Lambda(kh)}{\rho\mu_0}\left( \frac{k_xB_0}{k}\right)^2\Bigg]
\end{split}
\label{w_funk}
\end{equation}

\noindent and

\begin{equation}
\Lambda(kh)=\frac{e^{kh}(\mu_r+1)+e^{-kh}(\mu_r-1)}{e^{kh}(\mu_r+1)^2-e^{-kh}(\mu_r-1)^2} \; .\label{lambda}
\end{equation}

\noindent Here $\mathbf{B}_0=\mu_0\mathbf{H}_0$ is the induction of the applied magnetic field.

The essential differences of Eqs.\ (\ref{disp})--(\ref{lambda}) in comparison with the dispersion relation for the surface instability in 
a vertical magnetic  field \cite{Abou,Muller,Lange}
are the following: First, the term proportional to $B_0^2$ has the same sign as the terms related to surface tension and gravity. This
implies that a horizontal magnetic field alone cannot induce any instability. Second in the same term, the factor $k_x$ appears instead of 
$k$ in the case of the normal magnetic field. This allows one to introduce the effective field 

\begin{equation}
\mathbf{B}_{\text{eff}}=\frac{(\mathbf{k}\mathbf{B}_0)}{k^2}\mathbf{k}=B_0\,\cos(\theta)\frac{\mathbf{k}}{k} \; ,
\label{ef_field}
\end{equation}

\noindent where $\theta$ is the angle between $\mathbf{k}$ and $\mathbf{B}_0$.  Hence, the influence of a magnetic field with induction 
$\mathbf{B}_0$ on a wave propagating along an arbitrary direction $\mathbf{k}$
is equivalent to the influence of a field with induction $\mathbf{B}_{\text{eff}}$, which is parallel to $\mathbf{k}$.

Equation (\ref{disp})  has to be satisfied for all times which implies that each term of the sum equals to zero. 
Using the relations between $\zeta_n$ with positive and negative numbers (\ref{reality}), one gets the set of equations

\begin{subequations}
\label{set}
\begin{align}
W_0\zeta_0-a\zeta_1^*-a\zeta_1&=0,\quad \alpha=0 \; , \label{harm}\\
W_0\zeta_0-a\zeta_0^*-a\zeta_1&=0,\quad \alpha=1/2 \; , \label{subh}\\
W_n\zeta_n-a\zeta_{n-1}-a\zeta_{n+1}&=0,\quad n=1,\cdots \infty \; .\label{any_n}
\end{align}
\end{subequations}

\noindent A cutoff at $n=N$ (in the present work $N=100$) leads to a self-consistent equation for 
the acceleration amplitude $a$ \cite{Chen,Chen1},
                                                                                
\begin{equation}
a=|F(a,k,\omega,B_{\text{eff}},\nu,\sigma,\rho,\mu_r)| \; ,
\label{s-const}
\end{equation}

\noindent where $F$ is a complex function expressed in terms of  continued fractions. Equation (\ref{s-const}) can be solved 
numerically and gives the dependence of $a$ on $k$ at fixed parameters. The critical values of the acceleration amplitude 
$a_c$ and the wave number $k_c$ correspond to an absolute minimum of the curve $a(k)$ at zero growth rate ($s=0$).

\section{\label{results} Results and discussion}

\begin{figure}[bp]
    \includegraphics[scale=0.45]{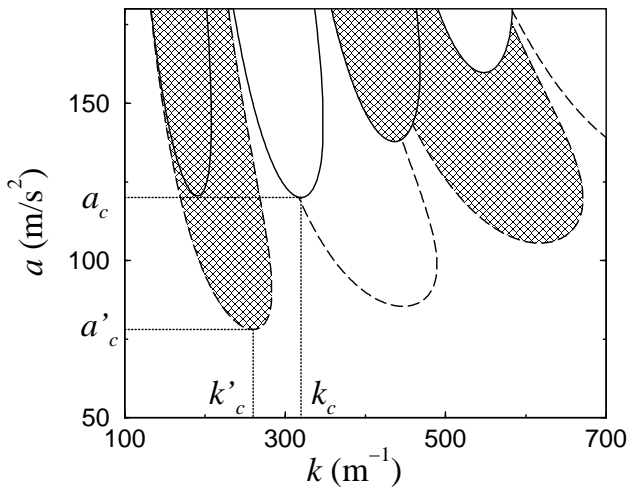}
    \caption{Neutral stability curves for the excitation frequency $f=10$ Hz and layer depth $h=2$ mm.    
Dashed (solid) lines correspond to $B_{\text{eff}}= 0$ ($B_{cR}$). Filled and unfilled tongues represent regions of subharmonic and harmonic 
responses. $a_c$ and $k_c$ ($a$'$_c$ and $k$'$_c$) are the critical acceleration amplitude and critical wave number  for
$B_{\text{eff}}=B_{cR}$ ($B_{\text{eff}}=0$). 
   The parameters of the fluid are $\nu=10^{-4}$ m$^2$/s, $\sigma= 0.0265$ N/m, $\rho = 1020$ kg/m$^3$, $\mu_r=1.85$, and $B_{cR}=17.28$ 
mT.}
  \label{tongues}
\end{figure}

In the following the effective field is given in units of the critical induction $B_{cR}$ for the Rosensweig instability 
on an infinitely deep layer of ferrofluid \cite{Rosensweig}. Figure 
\ref{tongues} presents marginal stability curves for a viscous ferrofluid at low frequency. The dependence of acceleration amplitude  on 
the wave number for $s=0$ divides the phase space into regions, where the surface of the ferrofluid is stable or unstable with respect to 
parametrically driven standing waves. The principal data which can be extracted are the critical acceleration amplitude, the wave number, 
and the number of the tongue to which they belong. The number of a tongue $l$ 
(from left to right) is the order of response: the basic wave frequency related to the $l$-th tongue is $\Omega=l\omega/2$. The 
odd and even tongues are the regions, where either a subharmonic or a  harmonic instability develops. It can be seen that all
tongues shift towards  smaller wave numbers under the influence of an applied magnetic field. In the case presented in Fig.\ \ref{tongues}
the field also causes a transition from  a subharmonic to a harmonic response. 

\begin{figure}[btp]
    \includegraphics[scale=0.45]{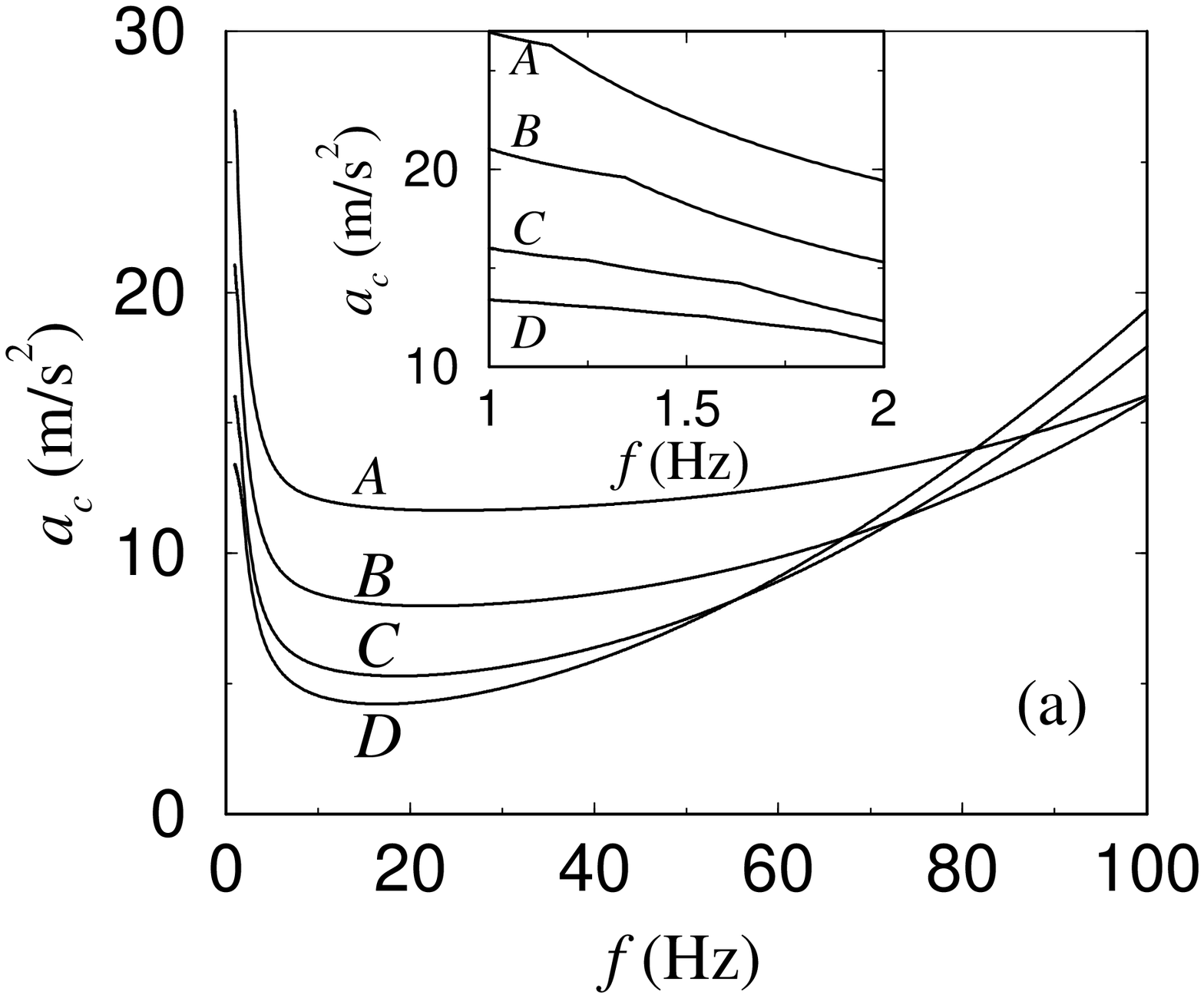}
    \includegraphics[scale=0.45]{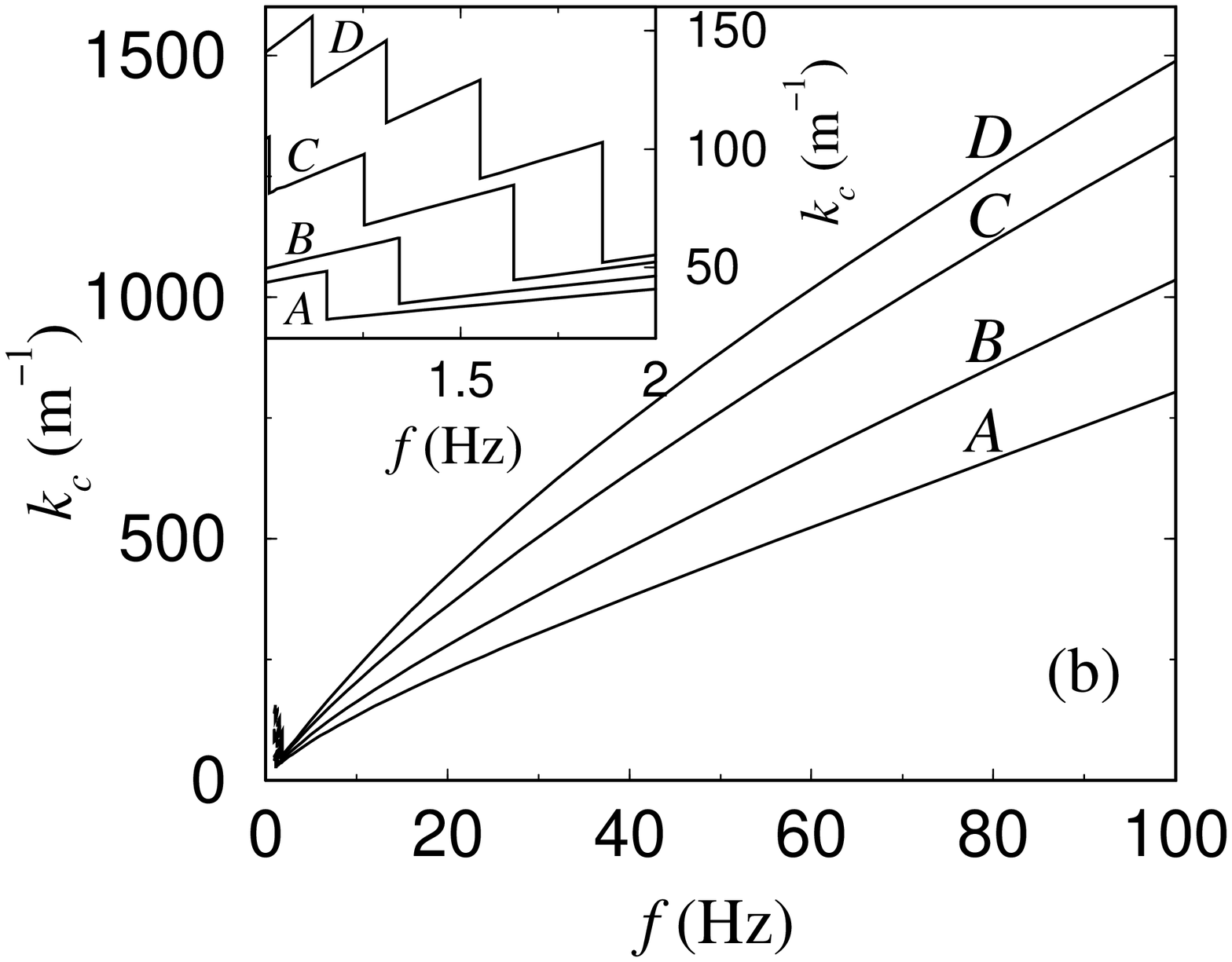}
    \caption{Frequency dependencies of  the critical acceleration amplitude $a_c$ (a) and  the critical
    wave number $k_c$ (b) for $h=2$ mm. Inserts: low-frequency behavior of the quantities. The effective
    field is $B=1.5 B_{cR}$ (curve {\it A}), $B_{cR}$ ({\it B}), $0.5 B_{cR}$ ({\it C}), 0 ({\it D}).
    The fluid parameters are  $\mu_r=1.85$, $\sigma=0.0265$  N/m, $\nu=5.88\times 10^{-6}$ m$^2$/s,
    $\rho =1020$ kg/m$^3$, and $B_{cR}=17.28$ mT.}
  \label{freq}
\end{figure}

The dependencies of the critical acceleration amplitude and the critical wave number on the
{\em excitation frequency} $f=\omega/2\pi$ are presented in Fig.\ \ref{freq}.
In the high frequency region, the instability is subharmonic, i.e. $l=1$. 
In the small frequency region bicritical points appear, which is why the
dependence of  $a_c(f)$ is not smooth and the  dependence  of $k_c(f)$ is discontinuous
(see inserts in Fig.\ \ref{freq}). Bicritical points are those points in the parameter space,
where the absolute minima of $a(k)$ is equal to the local minima of two neighbouring tongues.
In the zero field case for instance, such an overlap happens between the first
subharmonic tongue ($l=1$) and the first harmonic tongue ($l=2$) at $f_{\text{bc},1}\simeq 1.86$
Hz. For frequencies below $f_{\text{bc},1}$, the tongue of $2$nd order gives
the critical acceleration until the next bicritical point at $f_{\text{bc},2}\simeq 1.55$ Hz.
Below $f_{\text{bc},2}$ the tongue of $3$rd order gives there the lowest threshold until the
third bicritical point and so on. Thus with decreasing frequency subsequent transitions from
a tongue of the order $l$ to a tongue of the order $l+1$ occur.

\begin{figure}[btp]
    \includegraphics[scale=0.45]{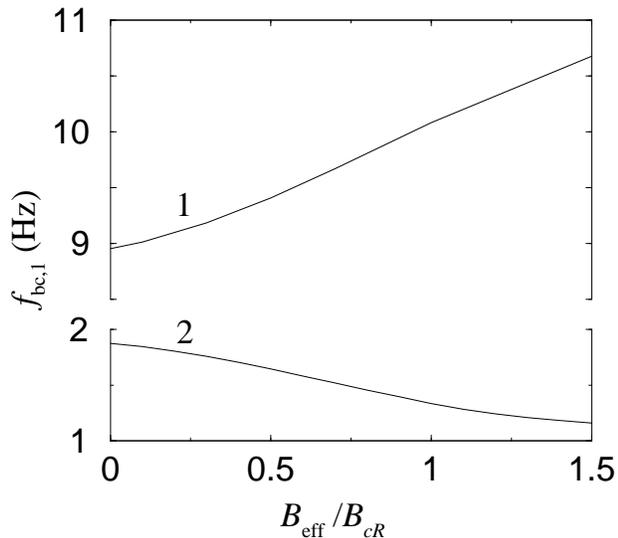}
    \caption{Frequency of the first bicritical point $f_{\text{bc},1}$ versus the effective magnetic
    induction. For a high-viscous fluid ($\nu=10^{-4}$ m$^2$/s, curve 1) an increasing induction
    increases $f_{\text{bc},1}$, whereas for a low-viscous fluid
    ($\nu=5.88\times 10^{-6}$ m$^2$/s, curve 2) $f_{\text{bc},1}$ is decreasing.}
  \label{bicrit}
\end{figure}

It is seen in the insert in Fig.\ \ref{freq}(b) that the magnetic field shifts the bicritical points.
If the fluid viscosity is low (as in Fig.\ \ref{freq}), then the magnetic field decreases the
frequencies of the transitions. For higher viscosities, e.g., $\nu=10^{-4}$ m$^2$/s, the bicritical
points are shifted towards higher frequencies. This is exemplarily shown in Fig.\ \ref{bicrit} for
the frequency of the first bicritical point, $f_{\text{bc},1}$, for a high-viscous fluid (curve 1)
and a low-viscous fluid (curve 2). For frequencies above (below) the curves the response of the
system is subharmonic (harmonic). Thus for a high-viscous fluid and an excitation frequency of $f=10$ Hz 
at zero applied field a subharmonic response ($\Omega=\omega/2$) is expected,
whereas at $B_{\text{eff}}=B_{cR}$ a harmonic response ($\Omega=\omega$) precedes.
For ordinary (nonmagnetic) fluids bicritical points were also
found in the case of both low-viscous fluids \cite{Muller1} and high-viscous fluids \cite{Cerda}.

A feature of the dependence of $a_c(f)$ [Fig.\ \ref{freq}(a)] is the appearance of a pronounced minimum for all tested 
fields. To explain this behavior, one has to recall that viscous damping is the reason for the finite (nonzero) value of the critical 
acceleration amplitude. The damping occurs due to the stresses in the  bottom layer and in the bulk fluid. The dimensionless 
quantity $d=k_ch$ can be used to determine which part of the damping is predominant. For a shallow fluid layer the inequality $d\lesssim 1$
holds and therefore the damping in the  bottom layer is predominant (first regime). For a deep layer, where the relation $d\gg 1$ is fulfilled,
the damping in the bottom layer is of no importance
and the damping in the bulk fluid is predominant (second regime).

The {\em first regime} occurs at the low frequency region in Fig.\ \ref{freq}, where $k_c\lesssim 500$ m$^{-1}$. Inside this region, as 
the frequency increases the effective depth $d$ of the fluid is increasing, too. Therefore the viscous stress in the  bottom layer
 is weakening and consequently the critical acceleration amplitude decreases. The {\em second damping regime} is typical for waves 
 with $k_c$ above 500 m$^{-1}$. Since dissipation in the bulk fluid is proportional to $k_c^2$ \cite{LL}, the damping becomes stronger and as a result the 
critical acceleration amplitude increases with frequency. The transition from the first to the second damping regime leads to the 
nonmonotonic dependence of the critical acceleration amplitude on the excitation frequency.

The most pronounced effect caused by the {\em magnetic field} is the decrease of the critical wave number [cf. curves \textit{D--A} in 
Fig.\ \ref{freq}(b)] as already observed in \cite{Zelazo}.
The reduction of the critical wave number affects the critical acceleration amplitude. This influence is different 
at low and high frequencies due to the following reason. The decrease of $k_c$ results in (i) the reduction of the effective fluid depth 
$d$, which intensifies the stress in the  bottom layer and (ii) the weakening of the dissipation in the bulk fluid. 
Therefore in the case of the first damping regime the viscous dissipation is intensified, whereas in the case of the second regime it is 
reduced by the magnetic field. 

In the {\em low frequency} region ($d=k_ch\lesssim 1$), the first damping regime occurs independent of the magnetic field. Therefore 
the critical acceleration amplitude increases with the growth of the effective field. 

At {\em high frequencies} the dependence of the critical acceleration amplitude on the effective field is more intricate. 
\begin{figure}[tbp]
    \includegraphics[scale=0.45]{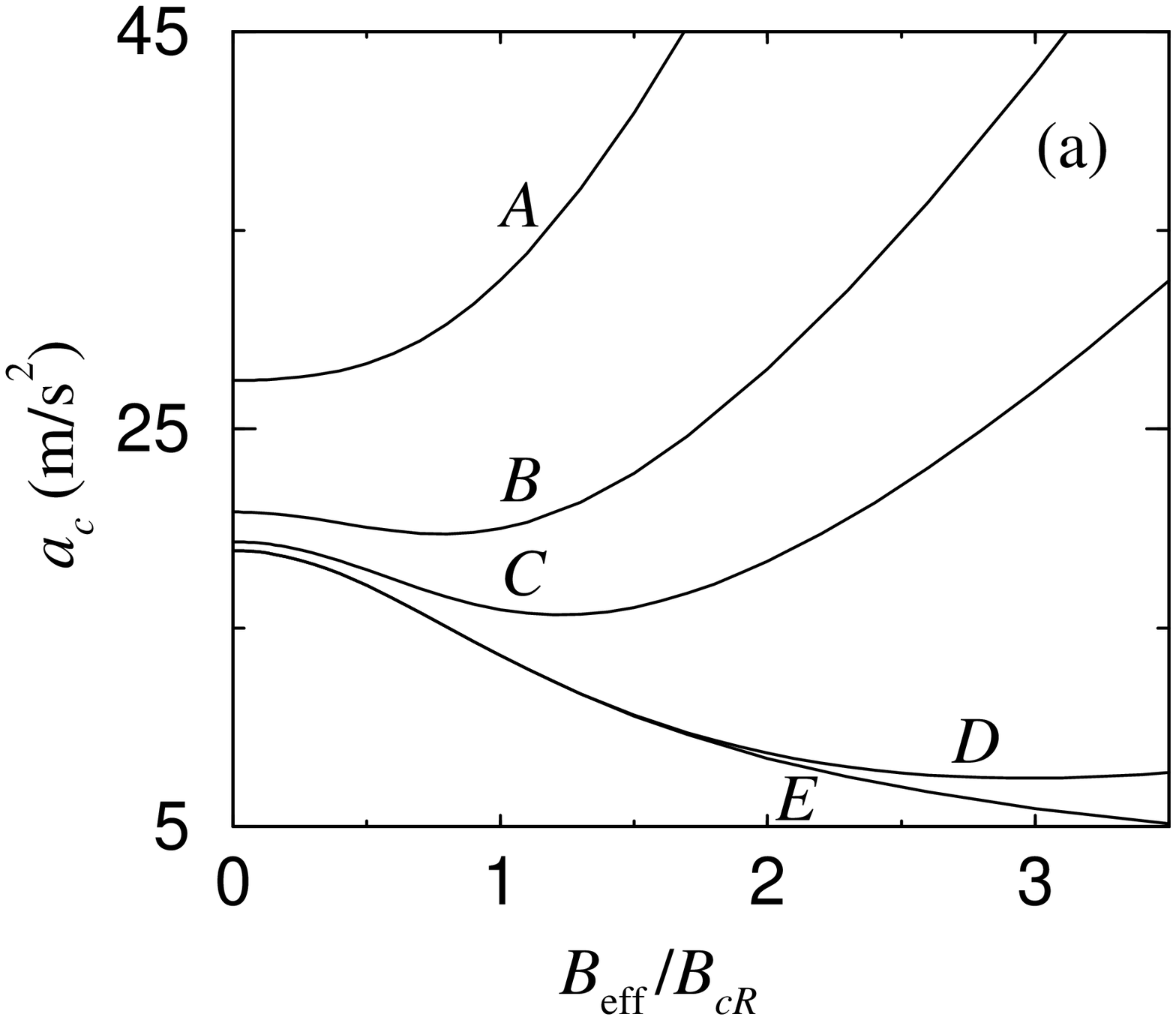}
    \includegraphics[scale=0.45]{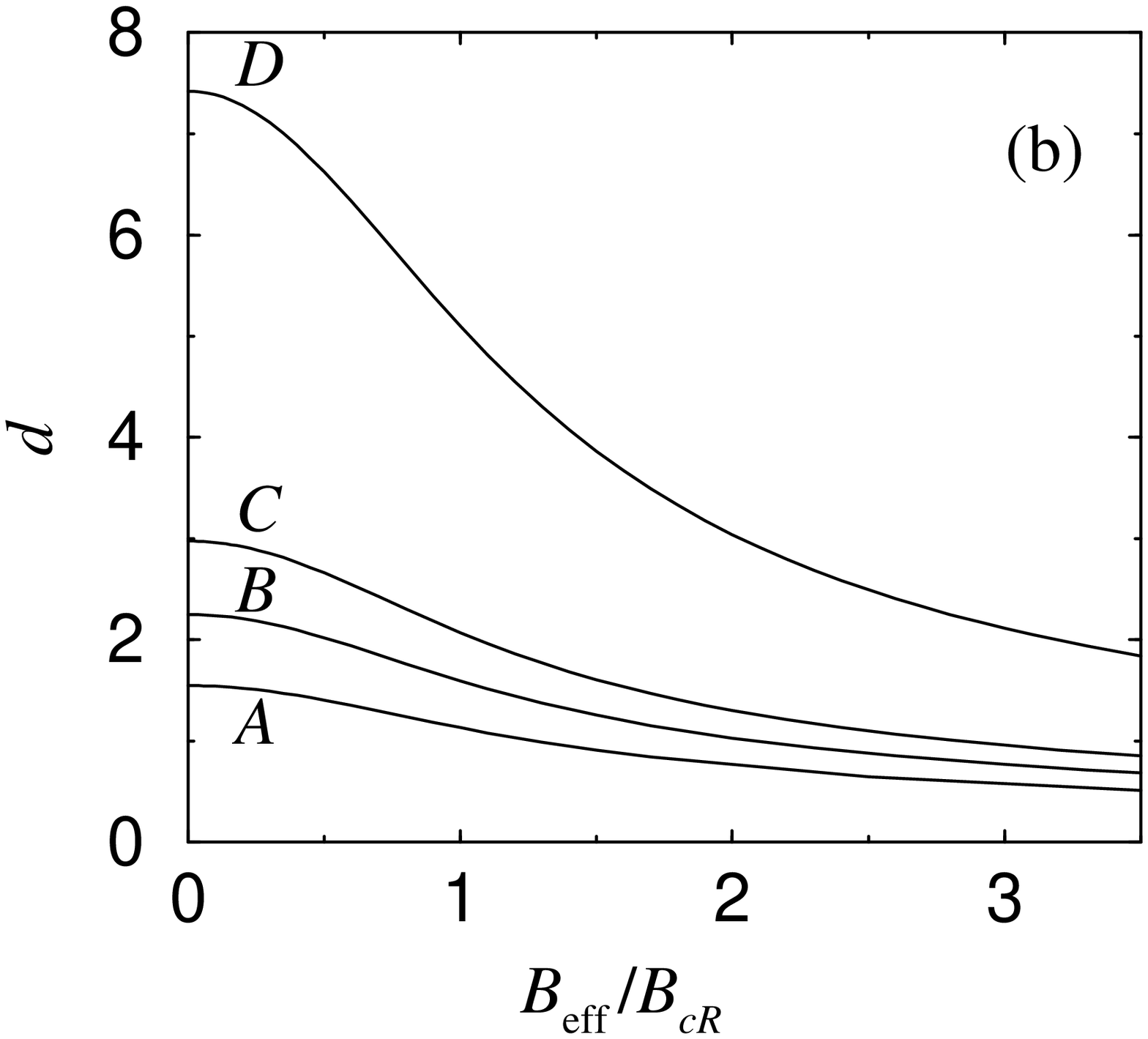}
    \caption{Critical acceleration amplitude $a_c$ (a) and the dimensionless depth $d=k_ch$ of the fluid (b) versus the induction of 
magnetic field for 
$f=100$ Hz.  The layer depth is 1 mm (curve {\it A}), 1.5 mm ({\it B}), 2 mm ({\it C}), 5 mm ({\it D}), and the curve ({\it E}) is 
obtained for an infinitely deep layer. The fluid parameters are those of Fig.\ \ref{freq}.}
  \label{field}
\end{figure}
In Fig.~\ref{field}(a) the dependence of $a_c(B_{\text{eff}})$ is presented for a frequency $f=100$ Hz. It is seen that  for an infinitely 
deep fluid the critical acceleration 
amplitude decreases monotonicly with the increase of the magnetic field. In the case of a finite depth of the layer, there are two 
different forms of the 
dependence of $a_c$ 
on $B_{\text{eff}}$. In the case of $h\ge 1.14$ mm  a minimum in the critical acceleration amplitude  is observed. It implies that a 
moderate magnetic 
field lowers the threshold value of the acceleration amplitude whereas a strong magnetic field stabilizes the surface, i.e., the critical 
acceleration amplitude is increasing. With the decrease of the layer depth [from curve {\it D} to {\it A} in Fig.\ \ref{field}(a)], the 
observed minimum shifts to the lower fields and becomes less pronounced. At $h\le 1.13$ mm the critical acceleration amplitude 
monotonicly increases with the effective field. 

To explain these qualitative changes in the dependence of $a_c(B_{\text{eff}})$ it is useful to consider $d(B_{\text{eff}})$ [see Fig.\ 
\ref{field}(b)]. The analysis shows that in the region $d\gg 1$ (the second damping regime) the decrease of the critical wave number 
caused by the magnetic field leads to a weakening of the viscous damping. Therefore the critical acceleration amplitude is decreasing with 
the increase of the field until the effective depth $d$ of the fluid becomes of the order of unity. A further decrease of $d$ caused by 
the magnetic field increases the viscous damping in the  bottom layer and consequently the critical acceleration amplitude. Thus, the transition 
from the second damping regime to the first one results in the nonmonotonic dependence of $a_c(B_{\text{eff}})$ [Fig.\ \ref{field}(a)]. 
The effective field, where the transition occurs, is decreasing with decrease of $h$, and in the case of a very shallow fluid (here, $h\le 
1.13$ mm), the second damping regime is not observed at all.

The nonmonotonic dependence of $a_c(B_{\text{eff}})$ allows one to select the type of the linearly most unstable pattern  
(corresponding to a minimal $a_c$) by applying a horizontal magnetic field. When the induction $B_0$ of the applied field  is smaller than 
$B_{\text{eff}}^*$ [$a_c(B_{\text{eff}})$ is minimal at $B_{\text{eff}}^*$] Faraday waves with $\mathbf{k}||\mathbf{B}_0$ are favorable. 
In this case $B_{\text{eff}}=B_0$ [see Eq.\ (\ref{ef_field})], i.e., $B_{\text{eff}}$ has a maximal possible  value  at 
given $B_0$. It implies that the first unstable pattern are rolls perpendicular to the magnetic field. If $B_0 \ge B_{\text{eff}}^*$ then 
the angle between the favorable wave vector of perturbation and the applied magnetic field satisfies the relation $\cos\theta=\pm 
B_{\text{eff}}^*/B_0$. Thus, for $B_0>B_{\text{eff}}^*\ne 0$ rolls with an angle $+\theta$ or an angle $-\theta$ or
a rhombic pattern as a superposition of both is most probable. If the layer depth is small enough 
$B_{\text{eff}}^*$ becomes zero [see curve {\it A} in Fig.\ \ref{field}(a)]. That is equivalent to $\cos\theta=0$, i.e., rolls parallel to 
the magnetic field are expected.

The parametrical generation of surface waves by a horizontal alternating magnetic field was observed in  \cite{Cebers,Bacri}. 
In these studies the driving force is the magnetic field and it is anisotropic. Only perturbations with a wave vector parallel to the 
field can be unstable in the linear approximation. Therefore the first unstable pattern observed in the experiments were always rolls
perpendicular to the field. In \cite{Bacri} a supercritical transition 
was observed from rolls to a rectangular pattern. This transition is caused by nonlinear interactions of the different modes. In the 
present study there is no anisotropic driving and therefore Faraday waves along an arbitrary direction can be excited. As a consequence, 
different patterns can be generated (rolls along an arbitrary direction or a rhombic pattern as discussed above).

In the case of {\em low viscosity} the obtained results for the critical acceleration amplitude are in the very good agreement with an
approximation suggested by M\"uller {\it et al.} (Eq.\ (9) in \cite{Muller1} and Eq.\ (4.1) in \cite{Muller}). 
It should be noted that the influence of a {\em normal} magnetic field (studied in \cite{Muller,Bacri_EPJ}) and a {\em horizontal} field 
(studied in the present paper) on Faraday waves are different. The normal magnetic field increases the critical wave number of the Faraday 
waves \cite{Bacri_EPJ}, whereas the horizontal magnetic field 
decreases the wave number. Thus, the sensitivity of ferrofluids to magnetic fields allows one both to increase and  to decrease the 
critical wave number. In this way the relative importance of the stress in the  bottom layer and the dissipation in the 
bulk fluid can be changed. Therefore magnetic fields are a convenient way to control the stability of the surface.

The Faraday instability on nonmagnetic fluids with {\em high viscosity} has been investigated in  \cite{Tuckerman}. The authors 
suggest a scale for the acceleration based on an analogy between the Faraday instability and a periodically driven version of the  
Rayleigh-Taylor instability. They observe a data collapse in the range of the dissipation parameter $(\delta/h)^2$ from 0.1 to 0.3, where 
$\delta=\sqrt{\nu/\omega}$ is the dissipative length scale. In this parameter range our results are in a good agreement with the suggested 
scaling law. However, at low frequencies the scaling behavior suggested in \cite{Tuckerman} strongly underestimates $a_c$. Additionally,  
the remaining parameters neglected in \cite{Tuckerman} become essential at $k_ch\lesssim 1$. Therefore, the scaling propagated in  
\cite{Tuckerman} is of limited use  for ferrofluids in magnetic fields.

\begin{figure}[tbp]
    \includegraphics[scale=0.45]{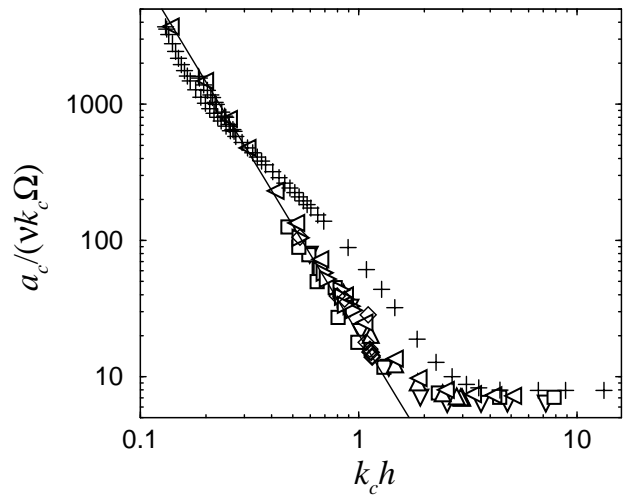}
    \caption{Scaled critical acceleration amplitude versus the dimensionless depth of the layer. 
Data are plotted for seven different sets with various combinations of the fluid viscosity, excitation frequency, effective field, and the 
layer thickness (see Table \ref{parametry}). The remaining parameters are those of Fig.\ \ref{freq}.
}
  \label{scaling}
\end{figure}
\begin{table}
\caption{\label{parametry}Sets of the parameters for data plotted in Fig.\ \ref{scaling}. }
\begin{ruledtabular}
\begin{tabular}{c c c c c c}
Set & $\nu$ (m$^2$/s)& $f$ (Hz)& $B_{\text{eff}}$ ($B_{cR}$)& $h$ (mm)& Symbol \\ \hline
1 & $5.88\times 10^{-6}$& 5& 1& 0.5\ldots 600& +\\
2 &$10^{-3}$&10&0&5\ldots 100&$\nabla$\\
3 & $10^{-4}$&100&0&1\ldots 100&$\Box$\\
4&$5.88\times 10^{-6}$&100&0\ldots 3& 2& $\triangle$\\
5&$5.88\times 10^{-6}$&1\ldots 100&1& 5&{\Large $\triangleleft$}\\
6&$10^{-3}$&5\ldots 100& 1&2&{\Large $\triangleright$}\\
7& $10^{-6}\ldots10^{-3}$& 48&1&2&{\Large $\diamond$}\\
\end{tabular}
\end{ruledtabular}
\end{table}

The aim now is to reveal a {\em scaling law} for the threshold values of the Faraday instability. The behavior of the critical 
acceleration amplitude as a function of the parameters of the system is determined by the dimensionless depth $d$ of the fluid (see Figs.\ 
\ref{freq},\ \ref{field}). Therefore, it is natural to look for a scaling law of the form $\bar{a}_c= \bar{a}_c(d)$, where $\bar{a}_c$ is 
the scaled acceleration. As $h\rightarrow\infty$  the required function should approach an asymptotic value. Thus, the appropriate scale 
for the acceleration amplitude is the value $a_{c,\infty}$ of the critical acceleration  for an infinitely deep fluid. The quantity 
$k_c\Omega\nu$ is a reasonable estimation for $a_{c,\infty}$ \cite{Tuckerman}. Introducing this  scale for the acceleration amplitude 
$\bar{a}_c=a_c/(\nu\Omega k_c)$, it is possible to map all the above presented  dependencies in a single plot as it has been done in Fig.\ 
\ref{scaling}.

Figure \ref{scaling} presents the dimensionless amplitude as a function of the effective depth $d=k_ch$ of the fluid. The latter can be 
varied by means of changing the physical depth of the layer (pluses, down triangles, and squares in Fig.\ \ref{scaling}), varying the 
applied magnetic field (up triangles), excitation frequency (left and right triangles), and the viscosity of the fluid (diamonds). It is 
seen that for a wide range of parameters the dependencies of $\bar{a}_c$ related to different fluids and varying parameters are in a 
rather good agreement with each other. The only deviation from the common behavior appears for a low frequency (pluses) at intermediate 
$d$. For  $d$ up to unity  the dimensionless acceleration amplitude can be reasonable approximated by $\bar{a}_c\approx 21 d^{-2.6}$ 
(solid line in Fig.\ \ref{scaling}). 
For $d\gg 1$ an estimation $\bar{a}_{c,\infty}\approx 7$ approximates the exact results with a maximal error of about 12 \%. It is worth 
noting 
that the dependencies of the critical acceleration amplitude on the surface tension and the fluid density follow the same common behavior 
as shown in Fig.\ \ref{scaling}.

\section{\label{conc} Conclusion} 

The linear analysis of the Faraday instability on a viscous ferrofluid subjected to a horizontal magnetic field has been performed. 
A horizontally unbounded ferrofluid layer of a finite depth  has been considered. The dependencies of the critical acceleration amplitude 
$a_c$ and the critical wave vector on the excitation frequency $f$ and the induction $B_{\text{eff}}$ of the  magnetic field   have been 
obtained  for different depths of the layer 
in a wide range of fluid viscosities. The regions have been found, where  the viscous stress either in the  bottom layer 
(the 
first regime) or in the bulk fluid (second regime) are predominant. A transition from the second damping regime to the first one can 
be caused by decreasing the excitation frequency or by applying a horizontal magnetic field. The transition results in the nonmonotonic 
dependencies of $a_c(f)$ and $a_c(B_{\text{eff}})$.  It is shown that one can select the first unstable pattern of Faraday waves by the
appropriate choice of the depth of the ferrofluid and the induction of the magnetic field. A scaling law is suggested, which describes the 
behavior of the system in a wide range of the system parameters.

\begin{acknowledgments}
This work was supported by the Deutsche Forschungsgemeinschaft under Grant
No. LA 1182/2.
\end{acknowledgments}

\bibliography{art}

\begin{thebibliography}{32}
\expandafter\ifx\csname natexlab\endcsname\relax\def\natexlab#1{#1}\fi
\expandafter\ifx\csname bibnamefont\endcsname\relax
  \def\bibnamefont#1{#1}\fi
\expandafter\ifx\csname bibfnamefont\endcsname\relax
  \def\bibfnamefont#1{#1}\fi
\expandafter\ifx\csname citenamefont\endcsname\relax
  \def\citenamefont#1{#1}\fi
\expandafter\ifx\csname url\endcsname\relax
  \def\url#1{\texttt{#1}}\fi
\expandafter\ifx\csname urlprefix\endcsname\relax\def\urlprefix{URL }\fi
\providecommand{\bibinfo}[2]{#2}
\providecommand{\eprint}[2][]{\url{#2}}

\bibitem[{\citenamefont{Faraday}(1831)}]{Faraday}
\bibinfo{author}{\bibfnamefont{M.}~\bibnamefont{Faraday}},
  \bibinfo{journal}{Philos.\ Trans.\ R.\ Soc.\ London}
  \textbf{\bibinfo{volume}{52}}, \bibinfo{pages}{319} (\bibinfo{year}{1831}).

\bibitem[{\citenamefont{{M\"uller} et~al.}(1997)\citenamefont{{M\"uller},
  Wittmer, Wagner, Albers, and Knorr}}]{Muller1}
\bibinfo{author}{\bibfnamefont{H.~W.} \bibnamefont{{M\"uller}}},
  \bibinfo{author}{\bibfnamefont{H.}~\bibnamefont{Wittmer}},
  \bibinfo{author}{\bibfnamefont{C.}~\bibnamefont{Wagner}},
  \bibinfo{author}{\bibfnamefont{J.}~\bibnamefont{Albers}}, \bibnamefont{and}
  \bibinfo{author}{\bibfnamefont{K.}~\bibnamefont{Knorr}},
  \bibinfo{journal}{Phys.\ Rev.\ Lett.} \textbf{\bibinfo{volume}{78}},
  \bibinfo{pages}{2357} (\bibinfo{year}{1997}).

\bibitem[{\citenamefont{Edwards and Fauve}(1994)}]{Fauve}
\bibinfo{author}{\bibfnamefont{W.~S.} \bibnamefont{Edwards}} \bibnamefont{and}
  \bibinfo{author}{\bibfnamefont{S.}~\bibnamefont{Fauve}},
  \bibinfo{journal}{J.\ Fluid Mech.} \textbf{\bibinfo{volume}{278}},
  \bibinfo{pages}{123} (\bibinfo{year}{1994}).

\bibitem[{\citenamefont{Binks et~al.}(1997)\citenamefont{Binks, Westra, and
  van~de Water}}]{Binks}
\bibinfo{author}{\bibfnamefont{D.}~\bibnamefont{Binks}},
  \bibinfo{author}{\bibfnamefont{M.-T.} \bibnamefont{Westra}},
  \bibnamefont{and} \bibinfo{author}{\bibfnamefont{W.}~\bibnamefont{van~de
  Water}}, \bibinfo{journal}{Phys.\ Rev.\ Lett.} \textbf{\bibinfo{volume}{79}},
  \bibinfo{pages}{5010} (\bibinfo{year}{1997}).

\bibitem[{\citenamefont{Binks and van~de Water}(1997)}]{Binks1}
\bibinfo{author}{\bibfnamefont{D.}~\bibnamefont{Binks}} \bibnamefont{and}
  \bibinfo{author}{\bibfnamefont{W.}~\bibnamefont{van~de Water}},
  \bibinfo{journal}{Phys.\ Rev.\ Lett.} \textbf{\bibinfo{volume}{78}},
  \bibinfo{pages}{4043} (\bibinfo{year}{1997}).

\bibitem[{\citenamefont{{M\"uller}}(1993)}]{Muller93}
\bibinfo{author}{\bibfnamefont{H.~W.} \bibnamefont{{M\"uller}}},
  \bibinfo{journal}{Phys.\ Rev.\ Lett.} \textbf{\bibinfo{volume}{71}},
  \bibinfo{pages}{3287} (\bibinfo{year}{1993}).

\bibitem[{\citenamefont{Kudrolli et~al.}(1998)\citenamefont{Kudrolli, Pier, and
  Gollub}}]{Kudrolli}
\bibinfo{author}{\bibfnamefont{A.}~\bibnamefont{Kudrolli}},
  \bibinfo{author}{\bibfnamefont{B.}~\bibnamefont{Pier}}, \bibnamefont{and}
  \bibinfo{author}{\bibfnamefont{J.~P.} \bibnamefont{Gollub}},
  \bibinfo{journal}{Physica D} \textbf{\bibinfo{volume}{123}},
  \bibinfo{pages}{99} (\bibinfo{year}{1998}).

\bibitem[{\citenamefont{Arbell and Fineberg}(1998)}]{Arbell1}
\bibinfo{author}{\bibfnamefont{H.}~\bibnamefont{Arbell}} \bibnamefont{and}
  \bibinfo{author}{\bibfnamefont{J.}~\bibnamefont{Fineberg}},
  \bibinfo{journal}{Phys.\ Rev.\ Lett.} \textbf{\bibinfo{volume}{81}},
  \bibinfo{pages}{4384} (\bibinfo{year}{1998}).

\bibitem[{\citenamefont{Arbell and Fineberg}(2000)}]{Arbell}
\bibinfo{author}{\bibfnamefont{H.}~\bibnamefont{Arbell}} \bibnamefont{and}
  \bibinfo{author}{\bibfnamefont{J.}~\bibnamefont{Fineberg}},
  \bibinfo{journal}{Phys.\ Rev.\ Lett.} \textbf{\bibinfo{volume}{84}},
  \bibinfo{pages}{654} (\bibinfo{year}{2000}).

\bibitem[{\citenamefont{Kumar and Tuckerman}(1994)}]{Kumar}
\bibinfo{author}{\bibfnamefont{K.}~\bibnamefont{Kumar}} \bibnamefont{and}
  \bibinfo{author}{\bibfnamefont{L.}~\bibnamefont{Tuckerman}},
  \bibinfo{journal}{J.\ Fluid Mech.} \textbf{\bibinfo{volume}{279}},
  \bibinfo{pages}{49} (\bibinfo{year}{1994}).

\bibitem[{\citenamefont{Bechhoefer et~al.}(1995)\citenamefont{Bechhoefer, Ego,
  Manneville, and Johnson}}]{Bechhoefer}
\bibinfo{author}{\bibfnamefont{J.}~\bibnamefont{Bechhoefer}},
  \bibinfo{author}{\bibfnamefont{V.}~\bibnamefont{Ego}},
  \bibinfo{author}{\bibfnamefont{S.}~\bibnamefont{Manneville}},
  \bibnamefont{and} \bibinfo{author}{\bibfnamefont{B.}~\bibnamefont{Johnson}},
  \bibinfo{journal}{J.\ Fluid Mech.} \textbf{\bibinfo{volume}{288}},
  \bibinfo{pages}{325} (\bibinfo{year}{1995}).

\bibitem[{\citenamefont{Weizhong and Rongjue}(1998)}]{Weizhong}
\bibinfo{author}{\bibfnamefont{C.}~\bibnamefont{Weizhong}} \bibnamefont{and}
  \bibinfo{author}{\bibfnamefont{W.}~\bibnamefont{Rongjue}},
  \bibinfo{journal}{Phys.\ Rev.\ E} \textbf{\bibinfo{volume}{57}},
  \bibinfo{pages}{4350} (\bibinfo{year}{1998}).

\bibitem[{\citenamefont{Cerda and Tirapegui}(1997)}]{Cerda}
\bibinfo{author}{\bibfnamefont{E.}~\bibnamefont{Cerda}} \bibnamefont{and}
  \bibinfo{author}{\bibfnamefont{E.}~\bibnamefont{Tirapegui}},
  \bibinfo{journal}{Phys.\ Rev.\ Lett.} \textbf{\bibinfo{volume}{78}},
  \bibinfo{pages}{859} (\bibinfo{year}{1997}).

\bibitem[{\citenamefont{Lioubashevski et~al.}(1997)\citenamefont{Lioubashevski,
  Fineberg, and Tuckerman}}]{Tuckerman}
\bibinfo{author}{\bibfnamefont{O.}~\bibnamefont{Lioubashevski}},
  \bibinfo{author}{\bibfnamefont{J.}~\bibnamefont{Fineberg}}, \bibnamefont{and}
  \bibinfo{author}{\bibfnamefont{L.~S.} \bibnamefont{Tuckerman}},
  \bibinfo{journal}{Phys.\ Rev.\ E} \textbf{\bibinfo{volume}{55}},
  \bibinfo{pages}{R3832} (\bibinfo{year}{1997}).

\bibitem[{\citenamefont{Kumar}(2000)}]{Kumar1}
\bibinfo{author}{\bibfnamefont{S.}~\bibnamefont{Kumar}},
  \bibinfo{journal}{Phys.\ Rev.\ E} \textbf{\bibinfo{volume}{62}},
  \bibinfo{pages}{1416} (\bibinfo{year}{2000}).

\bibitem[{\citenamefont{Milner}(1991)}]{Milner}
\bibinfo{author}{\bibfnamefont{S.~T.} \bibnamefont{Milner}},
  \bibinfo{journal}{J.\ Fluid Mech.} \textbf{\bibinfo{volume}{225}},
  \bibinfo{pages}{81} (\bibinfo{year}{1991}).

\bibitem[{\citenamefont{Zhang and {Vi\~nals}}(1995)}]{Zhang}
\bibinfo{author}{\bibfnamefont{W.}~\bibnamefont{Zhang}} \bibnamefont{and}
  \bibinfo{author}{\bibfnamefont{J.}~\bibnamefont{{Vi\~nals}}},
  \bibinfo{journal}{Phys.\ Rev.\ Lett.} \textbf{\bibinfo{volume}{74}},
  \bibinfo{pages}{690} (\bibinfo{year}{1995}).

\bibitem[{\citenamefont{Chen and {Vi\~nals}}(1999)}]{Chen1}
\bibinfo{author}{\bibfnamefont{P.}~\bibnamefont{Chen}} \bibnamefont{and}
  \bibinfo{author}{\bibfnamefont{J.}~\bibnamefont{{Vi\~nals}}},
  \bibinfo{journal}{Phys.\ Rev.\ E} \textbf{\bibinfo{volume}{60}},
  \bibinfo{pages}{559} (\bibinfo{year}{1999}).

\bibitem[{\citenamefont{Chen and {Vi\~nals}}(1997)}]{Chen}
\bibinfo{author}{\bibfnamefont{P.}~\bibnamefont{Chen}} \bibnamefont{and}
  \bibinfo{author}{\bibfnamefont{J.}~\bibnamefont{{Vi\~nals}}},
  \bibinfo{journal}{Phys.\ Rev.\ Lett.} \textbf{\bibinfo{volume}{79}},
  \bibinfo{pages}{2670} (\bibinfo{year}{1997}).

\bibitem[{\citenamefont{Silber and Skeldon}(1999)}]{Silber}
\bibinfo{author}{\bibfnamefont{M.}~\bibnamefont{Silber}} \bibnamefont{and}
  \bibinfo{author}{\bibfnamefont{A.~C.} \bibnamefont{Skeldon}},
  \bibinfo{journal}{Phys.\ Rev.\ E} \textbf{\bibinfo{volume}{59}},
  \bibinfo{pages}{5446} (\bibinfo{year}{1999}).

\bibitem[{\citenamefont{Rosensweig}(1993)}]{Ferrohydrodynamics}
\bibinfo{author}{\bibfnamefont{R.~E.} \bibnamefont{Rosensweig}},
  \emph{\bibinfo{title}{Ferrohydrodynamics}} (\bibinfo{publisher}{Cambridge
  University Press, Cambridge}, \bibinfo{year}{1993}).

\bibitem[{\citenamefont{Cowley and Rosensweig}(1967)}]{Rosensweig}
\bibinfo{author}{\bibfnamefont{M.~D.} \bibnamefont{Cowley}} \bibnamefont{and}
  \bibinfo{author}{\bibfnamefont{R.~E.} \bibnamefont{Rosensweig}},
  \bibinfo{journal}{J.\ Fluid Mech.} \textbf{\bibinfo{volume}{30}},
  \bibinfo{pages}{671} (\bibinfo{year}{1967}).

\bibitem[{\citenamefont{Abou et~al.}(2000)\citenamefont{Abou, Wesfreid, and
  Roux}}]{Abou00}
\bibinfo{author}{\bibfnamefont{B.}~\bibnamefont{Abou}},
  \bibinfo{author}{\bibfnamefont{J.-E.} \bibnamefont{Wesfreid}},
  \bibnamefont{and} \bibinfo{author}{\bibfnamefont{S.}~\bibnamefont{Roux}},
  \bibinfo{journal}{J.\ Fluid Mech.} \textbf{\bibinfo{volume}{416}},
  \bibinfo{pages}{217} (\bibinfo{year}{2000}).

\bibitem[{\citenamefont{{M\"uller}}(1999)}]{Muller}
\bibinfo{author}{\bibfnamefont{H.~W.} \bibnamefont{{M\"uller}}},
  \bibinfo{journal}{Phys.\ Rev.\ E} \textbf{\bibinfo{volume}{58}},
  \bibinfo{pages}{6199} (\bibinfo{year}{1999}).

\bibitem[{\citenamefont{P\'etr\'elis et~al.}(2000)\citenamefont{P\'etr\'elis,
  Falcon, and Fauve}}]{Petrelis}
\bibinfo{author}{\bibfnamefont{F.}~\bibnamefont{P\'etr\'elis}},
  \bibinfo{author}{\bibfnamefont{{\protect\'E}.}~\bibnamefont{Falcon}},
  \bibnamefont{and} \bibinfo{author}{\bibfnamefont{S.}~\bibnamefont{Fauve}},
  \bibinfo{journal}{Eur.\ Phys.\ J.\ B} \textbf{\bibinfo{volume}{15}},
  \bibinfo{pages}{3} (\bibinfo{year}{2000}).

\bibitem[{\citenamefont{Browaeys et~al.}(1999)\citenamefont{Browaeys, Bacri,
  Flament, Neveu, and Perzynski}}]{Bacri_EPJ}
\bibinfo{author}{\bibfnamefont{J.}~\bibnamefont{Browaeys}},
  \bibinfo{author}{\bibfnamefont{J.-C.} \bibnamefont{Bacri}},
  \bibinfo{author}{\bibfnamefont{C.}~\bibnamefont{Flament}},
  \bibinfo{author}{\bibfnamefont{S.}~\bibnamefont{Neveu}}, \bibnamefont{and}
  \bibinfo{author}{\bibfnamefont{R.}~\bibnamefont{Perzynski}},
  \bibinfo{journal}{Eur. Phys.\ J. B} \textbf{\bibinfo{volume}{9}},
  \bibinfo{pages}{335} (\bibinfo{year}{1999}).

\bibitem[{\citenamefont{Cebers and Maiorov}(1989)}]{Cebers}
\bibinfo{author}{\bibfnamefont{A.}~\bibnamefont{Cebers}} \bibnamefont{and}
  \bibinfo{author}{\bibfnamefont{M.~M.} \bibnamefont{Maiorov}},
  \bibinfo{journal}{Magnintnaya Gidrodinamika} \textbf{\bibinfo{volume}{4}},
  \bibinfo{pages}{38} (\bibinfo{year}{1989}).

\bibitem[{\citenamefont{Bacri et~al.}(1994)\citenamefont{Bacri, Cebers,
  Dabadie, and Perzynski}}]{Bacri}
\bibinfo{author}{\bibfnamefont{J.-C.} \bibnamefont{Bacri}},
  \bibinfo{author}{\bibfnamefont{A.}~\bibnamefont{Cebers}},
  \bibinfo{author}{\bibfnamefont{J.-C.} \bibnamefont{Dabadie}},
  \bibnamefont{and}
  \bibinfo{author}{\bibfnamefont{R.}~\bibnamefont{Perzynski}},
  \bibinfo{journal}{Phys.\ Rev.\ E} \textbf{\bibinfo{volume}{50}},
  \bibinfo{pages}{2712} (\bibinfo{year}{1994}).

\bibitem[{\citenamefont{Zelazo and Melcher}(1969)}]{Zelazo}
\bibinfo{author}{\bibfnamefont{R.~E.} \bibnamefont{Zelazo}} \bibnamefont{and}
  \bibinfo{author}{\bibfnamefont{J.~R.} \bibnamefont{Melcher}},
  \bibinfo{journal}{J. Fluid Mech.} \textbf{\bibinfo{volume}{39}},
  \bibinfo{pages}{1} (\bibinfo{year}{1969}).

\bibitem[{\citenamefont{Abou et~al.}(1997)\citenamefont{Abou, de~Surgy, and
  Westfreid}}]{Abou}
\bibinfo{author}{\bibfnamefont{B.}~\bibnamefont{Abou}},
  \bibinfo{author}{\bibfnamefont{G.~N.} \bibnamefont{de~Surgy}},
  \bibnamefont{and} \bibinfo{author}{\bibfnamefont{J.~E.}
  \bibnamefont{Westfreid}}, \bibinfo{journal}{J. Phys. II}
  \textbf{\bibinfo{volume}{7}}, \bibinfo{pages}{1159} (\bibinfo{year}{1997}).

\bibitem[{\citenamefont{Lange et~al.}(2000)\citenamefont{Lange, Reimann, and
  Richter}}]{Lange}
\bibinfo{author}{\bibfnamefont{A.}~\bibnamefont{Lange}},
  \bibinfo{author}{\bibfnamefont{B.}~\bibnamefont{Reimann}}, \bibnamefont{and}
  \bibinfo{author}{\bibfnamefont{R.}~\bibnamefont{Richter}},
  \bibinfo{journal}{Phys.\ Rev.\ E} \textbf{\bibinfo{volume}{61}},
  \bibinfo{pages}{5528} (\bibinfo{year}{2000}).

\bibitem[{\citenamefont{Landau and Lifshitz}(1959)}]{LL}
\bibinfo{author}{\bibfnamefont{L.}~\bibnamefont{Landau}} \bibnamefont{and}
  \bibinfo{author}{\bibfnamefont{E.}~\bibnamefont{Lifshitz}},
  \emph{\bibinfo{title}{Fluid Mechanics}} (\bibinfo{publisher}{Pergamon, New
  York}, \bibinfo{year}{1959}).

\end{thebibliography}
\end{document}